\documentclass{aa}  

\usepackage[active]{srcltx}
\usepackage{graphicx}
\usepackage{amsmath}
\usepackage{amssymb}
\usepackage{aalongtable}
\usepackage{natbib}
\bibpunct{(}{)}{;}{a}{}{,} 

\usepackage{txfonts}
\defcitealias{khea12}{Paper~I}
\defcitealias{mwscat}{Paper~II}
\begin{document}

   \title{Global survey of star clusters in the Milky Way}
   \subtitle{III. 139 new open clusters at high Galactic latitudes}

   \author{S. Schmeja\inst{1}
          \and
          N.~V. Kharchenko\inst{1,2}
          \and
          A.~E. Piskunov\inst{1,3}
          \and
          S. R\"oser\inst{1}
          \and
          E. Schilbach\inst{1}
          \and
          D. Froebrich\inst{4}
          \and
          R.-D. Scholz\inst{5}
          }

   \institute{Astronomisches Rechen-Institut, Zentrum f\"ur Astronomie der
Universit\"at Heidelberg,
              M\"onchhofstr.~12-14, 69120 Heidelberg, Germany\\
              \email{sschmeja@ari.uni-heidelberg.de}
         \and
             Main Astronomical Observatory, 27 Academica Zabolotnogo Str., 03680
Kiev, Ukraine
         \and
             Institute of Astronomy of the Russian Academy of Sciences, 48
Pyatnitskaya Str., 109017 Moscow, Russia
         \and
             Centre for Astrophysics and Planetary Science, University of Kent,
Canterbury, CT2 7NH, United Kingdom
         \and
             Leibniz-Institut für Astrophysik Potsdam (AIP), An der Sternwarte
16, 14482 Potsdam, Germany
             }

   \date{Received xxx; accepted yyy}

  \abstract
   {An earlier analysis of the Milky Way Star Cluster (MWSC) catalogue revealed an apparent lack of old ($t \gtrsim 1$\,Gyr) open clusters in the solar neighbourhood ($d \lesssim 1$\,kpc).}
   {To fill this gap we undertook a search for hitherto unknown star clusters, assuming that the missing old clusters reside at high Galactic latitudes $|b|> 20 \degr$.}
{We were looking for stellar density enhancements using a star count algorithm on the 2MASS point source catalogue. To increase the contrast between potential clusters and the field, we applied filters in colour-magnitude space according to typical colour-magnitude diagrams of nearby old open clusters. The subsequent comparison with lists of known objects allowed us to select thus far unknown cluster candidates. For verification they were processed with the standard pipeline used within the MWSC survey for computing cluster membership probabilities and for determining structural, kinematic, and astrophysical parameters.}
   {In total we discovered 782 density enhancements, 522 of which were classified as real objects. Among them 139 are new open clusters with ages $8.3 < \log (t~\rm{[yr]}) < 9.7$, distances $d<3$\,kpc, and distances from the Galactic plane $0.3 < Z < 1$\,kpc. This new sample has increased the total number of known high latitude open clusters by about 150\%. Nevertheless, we still observe a lack of older nearby clusters up to 1 kpc from the Sun. This volume is expected to still contain about 60 unknown clusters that probably escaped our detection algorithm, which fails to detect sparse overdensities with large angular size.
   }
   {}


   \maketitle
%

\section{Introduction}

With this paper we continue to present the results of the Milky Way Star Cluster (MWSC) survey undertaken on the basis of the two all-sky catalogues, 2MASS \citep{cat2mass} and PPMXL \citep{ppmxl}. The MWSC survey was initiated a few years ago with the aim of building a comprehensive sample of Galactic star clusters with well-determined parameters, which is complete enough to enable an unbiased study of the content and evolution of the star clusters of our Galaxy. The first paper of this series \citep{khea12}, called hereafter \citetalias{khea12}, gave an introduction to the survey, explained the underlying motivation, provided a short review of similar studies, described the observational basis of the survey, the data processing pipeline, and presented preliminary results obtained in the second Galactic quadrant. The second paper \citep[][Paper~II]{mwscat} summarises the results of the full survey carried out for a compiled input list of 3784 known objects, covering the whole sky. It presents uniform structural, kinematic, and astrophysical data for 3006 open clusters, globular clusters, and compact associations.

\begin{figure}
\resizebox{\hsize}{!}{\includegraphics{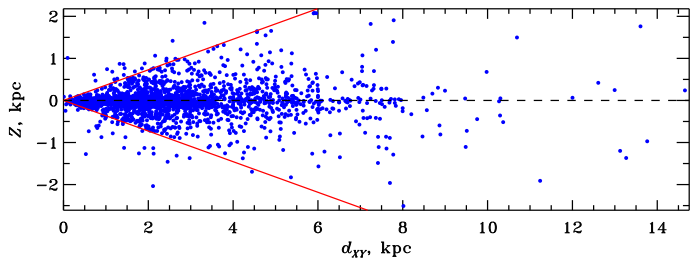}}
\caption{Distribution of Galactic open clusters from the MWSC survey in the plane $(Z,d_{XY})$. Solid lines show the limits corresponding to $b=\pm20\degr$. The dashed line marks the Galactic plane.}
\label{fig:zdxy}
\end{figure}

The first-look analysis of the MWSC data carried out in \citetalias{mwscat} has shown that the MWSC sample is complete up to a distance of $d=1.8$~kpc from the Sun for clusters of all ages except the older clusters ($\log (t~\rm{[yr]})>$9). Although this shortage concerns primarily the oldest clusters, the effect can be seen in the general distribution of all Galactic open clusters in Fig.~\ref{fig:zdxy}, where we show the cluster distribution in the plane $[Z,d_{XY}]$, with $Z$ the vertical distance from the Galactic plane, and $d_{XY}$ the distance from the Sun projected onto the Galactic plane. One can clearly see that at $d_{XY}\lesssim 2$ kpc the number of high-latitude clusters diminishes with decreasing $d_{XY}$.

The general lack of old open clusters has already been noted in the 1950s \citep[e.g.][]{oort58}. Since then, old clusters have been mainly discovered at distances $\gtrsim 1$\,kpc, resulting in a striking apparent absence of old clusters in the solar neighbourhood.

There are two main reasons that nearby old open clusters may have escaped previous searches, both based on their proximity.
\begin{enumerate}
\item Old open clusters show a larger scale height \citep{vdb+mccl80, froea10}, so in combination with small distances they may be located at higher Galactic latitudes, while systematic searches for open clusters have typically been restricted to areas close to the Galactic plane (e.g.\ \citealt{mercl05}: $|b| < 1 \degr$; \citealt{froeb07}: $|b| < 20 \degr$; \citealt{gluea10}: $|b| < 24 \degr$);
\item With a large angular extent (up to several degrees), they do not stand out prominently as overdensities from the field.
\end{enumerate}

The primary goal of this paper is to get a complete list of clusters within the MWSC survey. To reach this goal we expand previous searches of star clusters in 2MASS performed typically at $|b| \lesssim 20 \degr$ to higher Galactic latitudes.
This work can be considered as an extension of the search by \citet{froeb07}, which used the same data and a similar approach (without filters), but was restricted to the area $|b| < 20 \degr$.

In Section~\ref{sec:method} we describe the data set and our method of identifying clusters. The results are presented in Section~\ref{sec:results} and discussed in Section~\ref{sec:discuss}.

\section{Method}
\label{sec:method}

\subsection{Data}

Cluster candidates were identified as density enhancements in the Two Micron All Sky Survey (2MASS) point-source catalogue \citep{cat2mass}. The 2MASS survey provides the photometric basis of the MWSC survey with a uniformly calibrated photometry of the entire sky, complete down to $K_s \approx 14.3$~mag, depending on the position on the sky. We only considered sources that were detected in all three bands ($J$, $H$, $K_s$) with high quality ({{\it R\_flg} = 1, 2 or 3). We applied our search algorithm to the entire sky at Galactic latitudes $|b|>20 \degr$.

\subsection{Filtering the sample}\label{sec:filt} 

\begin{figure}
\resizebox{\hsize}{!}{\includegraphics{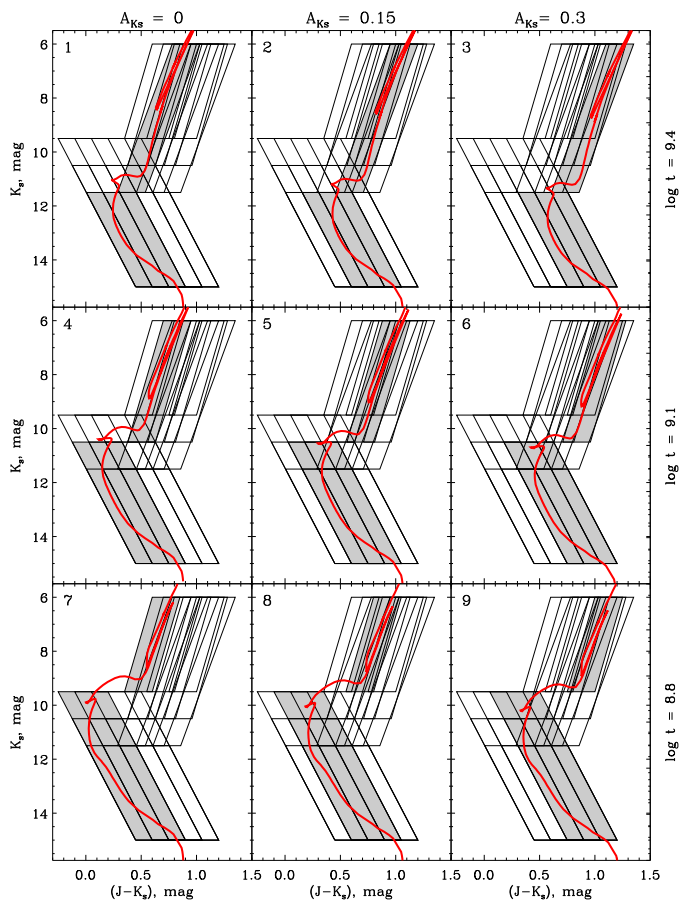}}
\caption{Nine overlapping filters used to reduce the contamination by field stars. 
The respective filter is highlighted in grey, the corresponding isochrone for the age and extinction at 1~kpc indicated
is shown as a red line. Similar filters were used for a distance of 0.5~kpc. The filter number is given in the upper left corner of each panel.
}
\label{fig:filters}
\end{figure}

Since nearby old clusters may not exhibit a significant overdensity in the plain 2MASS, we had to enhance the contrast between potential clusters and the field. Therefore, we used cuts in colour and magnitude according to typical colour-magnitude diagrams (CMDs) of clusters in different age, distance, and extinction bins. This is an approach comparable, albeit somewhat simpler, to what has been used to detect tidal tails of globular clusters \citep[e.g.][]{grillmair95,oden03}. We set up nine different filters to cover the colour-magnitude space expected for clusters with $0 \lesssim A_{K_s} \lesssim 0.3$~mag and $8.8 \lesssim \log (t~\rm{[yr]}) \lesssim 9.4$ at distances around 0.5 and 1~kpc (Fig.~\ref{fig:filters}). 
This filtering procedure reduces the number of sources in a field to between about 10 and 40 per cent. Figure~\ref{fig:density_map} illustrates the effect of the filtering: While no significant density enhancement can be detected in the unfiltered distribution, a density enhancement above the $4 \sigma$ level shows up after applying one of the filters. This feature is subsequently confirmed as an open cluster (\object{MWSC 5723}).

The filters are not designed to model a specific type of cluster, but to cover the parameter range in the CMD expected for clusters in the desired age and distance range, to reduce the contamination from unrelated background objects.
Since the filters are rather wide, overlap strongly and occupy a wide range, they also cover other parameter combinations,
in particular for smaller and larger distances.

\begin{figure}
\resizebox{0.975\hsize}{!}{\includegraphics[clip=]{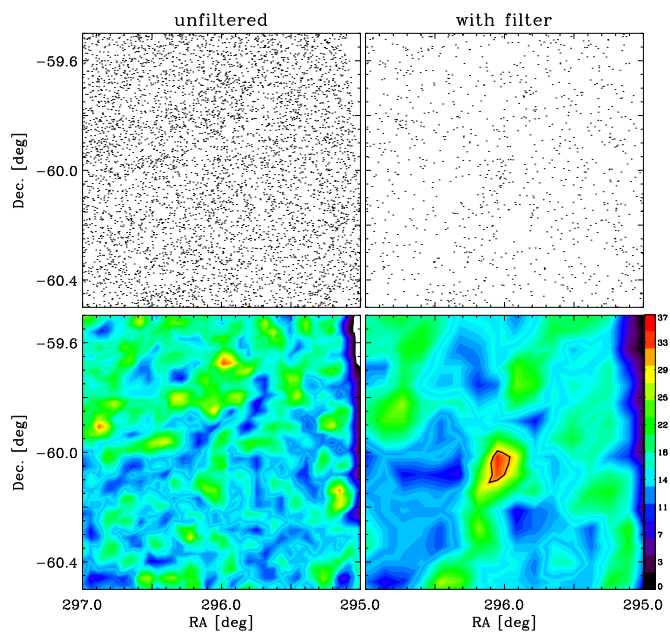}}
\caption{A $1 \degr \times 1 \degr$ field around the newly found cluster MWSC 5723: 2MASS point sources (upper row) and stellar density maps in number of stars per bin (lower row) for the unfiltered sample (left) and after applying one of the filters (right). The black line on the stellar density map indicates the 4~$\sigma$ contour.}
\label{fig:density_map}
\end{figure}

\subsection{Finding cluster candidates}

The filtered sample, together with the unfiltered catalogue, is then used as input for a cluster search algorithm based on star counts \citep[e.g.][]{carpenter95,lada+lada95,ivanov02,reyle+robin02}. This rather simple approach is nevertheless a very efficient way of creating stellar density maps and identifying density enhancements in a field, comparable to or better than more sophisticated approaches, such as the nearest neighbour density or the separation of minimum spanning trees \citep{schmeja11}. We use fields of $5 \degr \times 5 \degr$ in size. Every field is subdivided into a rectilinear grid of overlapping squares that are separated by half the side length of an individual square (the Nyquist spatial sampling interval). The size of the bins is chosen such that they contain 15 stars on average. This results in bins with side lengths between about 3 and 20~arcmin. All areas showing a density $\ge 4 \sigma$ above the average density of the field are considered potential clusters, if they contain at least ten sources. 
Tests showed that bins with a size that on average gives 15 stars per bin and an 
overdensity threshold of $4 \sigma$ are best suited to detecting clusters
 without missing  a signifiant number of clusters and picking up too many random density enhancements.
Density enhancements that by visual inspection could obviously not be Galactic stellar clusters (such as fragments of M31 or the Magellanic Clouds) were neglected. As a result we prepared a list of candidate clusters that contains the coordinates of the centres of the density enhancements and their sizes.

\subsection{Veryfing the candidates and determination of the cluster parameters}

To be certain that we do not re-discover already known objects, we checked if any candidate was in the MWSC input list, in the SIMBAD data base\footnote{http://simbad.u-strasbg.fr/simbad/}, and because many compact galaxies may appear as point sources in the 2MASS, in the list of galaxy clusters from the \citet{abell89} catalogue. The correctness of the preliminary choice of the candidate objects is supported by frequent coincidence of the candidates found with already known objects. The list of unidentified candidates, together with preliminary data on their positions and sizes, was processed with the MWSC pipeline for further checks for the construction of cluster membership and parameter determination.

The pipeline uses kinematic, photometric, and spatial information on stars in the candidate area and is described in more detail in \citetalias{khea12}. The main purpose of the pipeline is to clean a candidate from the fore- and background contamination using kinematic, photometric, and spatial criteria, to produce a list of probable members and to determine the basic cluster parameters in the case of success. The pipeline consists of iterative series of interactive checks of vector point diagram of proper motions, radial density profiles, magnitude-proper motion relation, and various colour-magnitude, two-colour, and $Q_{JHK_s}$-colour diagrams. As a theoretical basis, we use recent Padova stellar models of \citet{marigo08} and \citet{girardi08} with isochrones computed with the CMD2.2 on-line server\footnote{http://stev.oapd.inaf.it/cgi-bin/cmd}, whereas the pre-main sequence isochrones were computed by us from the models of \citet{siessea00} and then transformed to the $JHK_s$ photometric system using transformation tables provided by the Padova team with the \textit{dustyAGB07} database\footnote{http://stev.oapd.inaf.it/dustyAGB07/}. The membership probabilities of stars in the diagrams take data accuracy into account and are determined from the star location with respect to the reference sequences (represented either by isochrones or the average cluster proper motion), which themselves depend on the cluster parameters we want to find. This therefore requires an iterative approach, allowing us to successively improve both cluster membership and cluster parameters. The initial approximation was based on a visual inspection of the diagrams. As a rule, the process converges after a few iterations. Including spatial and kinematic criteria greatly helps reduce ambiguities in determining age, distance, and reddenning, which may arise if only photometric membership is considered. Details of this effect, called degeneracy, are described in detail in Paper I (Sec. 3.4.3).

The verification of the overdensities as clusters
is based on the most probable members only with $P_m> 61\%$ (deviating
from the reference by less than one \textit{rms}-error).
 If their distribution in the vector point diagram
of proper motions is more
compact than for the rest of the stars, and if they fit the critical
points of the isochrone (turn-off, red-giant branch), a candidate
is considered to be confirmed, and the most probable members
are used for computing the cluster parameters. Otherwise it is
rejected as a random clustering of field stars (asterism).
The verification by visual
inspection of the diagrams is supported by objective statistical
arguments. Applying a Fisher test to the identified clusters, we
find that the populations of the most probable cluster members
($P_m > 61 \%$) and of ``field'' stars ($P_m < 1\%$) have significantly different dispersions both in the vector point diagram (for 120, or 88\% of the clusters) and in the CMD (for all clusters). Figures~\ref{fig:atlp1} and \ref{fig:atlp2} show the atlas page of an exemplary cluster (MWSC 5224) with its spatial distribution, the radial density profile, the CMDs, and proper motion diagrams.

\begin{figure}
\resizebox{0.975\hsize}{!}{\includegraphics[clip=]{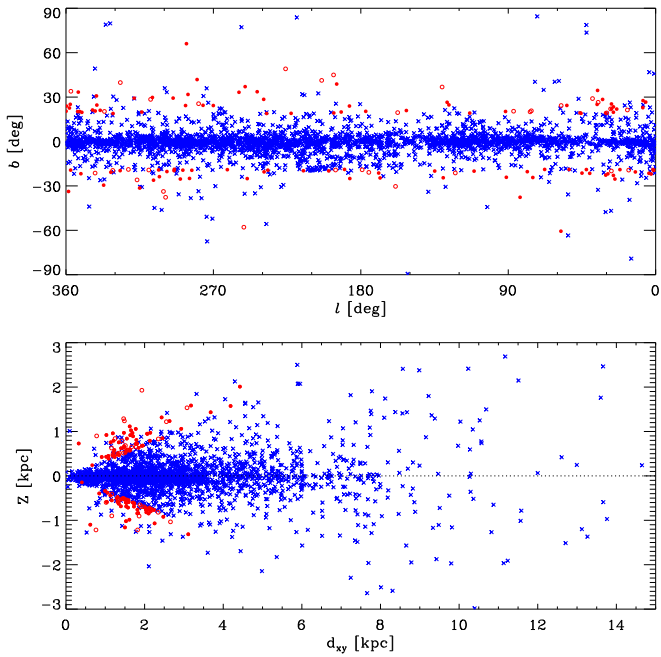}}
\caption{Distribution on the sky (upper panel) and in the plane $(Z,d_{XY})$ (lower panel) of known MWSC open clusters (blue crosses) and newly detected clusters (red circles: filled for clusters detected using the filters, open for clusters detected without filters).}
\label{fig:bl_map}
\end{figure}

\begin{table}[b]
\caption{Classification of star cluster candidates}
\label{tbl:sch_stat}
\begin{tabular}{lrrr}
\hline
\noalign{\smallskip}
Object&$b>20 \degr$&$b<-20 \degr$ &All \\
\hline
\noalign{\smallskip}
New clusters   & 74   &  65 & 139 \\
Known objects  & 206  & 179 & 385 \\
Asterisms      & 134  & 126 & 260 \\
\noalign{\smallskip}
Total          & 414  & 370 & 784  \\
\hline
\noalign{\smallskip}
\end{tabular}\\
\end{table}

\section{Results}
\label{sec:results}

The statistics of results of our cluster search is given in Table~\ref{tbl:sch_stat}, which shows the number of candidates, divided into three groups of objects: new real clusters, asterisms, and re-identified known stellar or galaxy clusters. About half of the candidates match known objects: 338 galaxy clusters, 33 globular and 6 Milky Way open clusters, and 8 clusters in the Large Magellanic Cloud. Comparing these statistics to the data present in the catalogues, we can estimate the efficiency of the applied search algorithm. At $|b|>20 \degr$ there are 49 Galactic globular clusters in the catalogue of \citet[edition 2010]{harris96}. This means that we were able to detect 67\% of the known globular clusters. The remaining globular clusters are too faint or too poorly represented in 2MASS to be detected.
There were 61 open clusters at $|b|>20$\degr\ in the MWSC catalogue prior to this work. Excluding associations, 
moving groups, embedded clusters, and cluster remnants from the sample, there are 18 clusters (called 'compact' here), of which we were able to identify six (NGC~188, NGC~2682, NGC~1662, NGC~1980, NGC~2632, and Blanco~1), corresponding to a detection rate of 10\% of all open clusters or 33\% of the compact clusters. 
According to the SIMBAD database, there are 26\,227 clusters of galaxies at $|b|>20 \degr$. For those our detection rate is of the order of 1\% (338). The detection rate of open clusters is therefore ten times higher than the detection rate of clusters of galaxies.

Out of the 139 new clusters, 104 were detected using the CMD filters described in Sec.~\ref{sec:filt}, 34 were only found without filters, and one was detected by both applying one of the filters and using the unfiltered field. Since we performed both a filtered and an unfiltered search, we found clusters outside the targeted age and distance limits implied by the filters.

\begin{figure}
\resizebox{\hsize}{!}{\includegraphics[clip=]{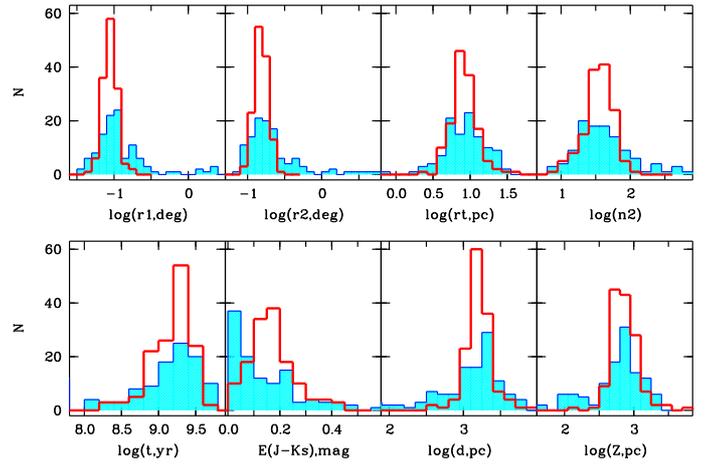}}
\caption{The distributions of the parameters of new clusters (open red histograms) and of MWSC open clusters at $|b|>20 \degr$ (blue filled histograms). The upper row compares the distributions of ``structural'' parameters. The bottom row gives the distributions of ``photometric'' parameters. See text for an explanation of the definitions.
}
\label{fig:cmprm}
\end{figure}

In Fig.~\ref{fig:bl_map} we show the distribution of the newly discovered clusters on the sky, together with the previosuly known open clusters from the MWSC survey. The majority of the confirmed clusters are located within $|b|\lesssim 30^\circ$, though a few open clusters were found up to $|b|\approx 60^\circ$. However, most of the high-latitude candidates turned out to be galaxy clusters.

In Fig.~\ref{fig:cmprm} we compare the distributions of the parameters of newly detected clusters and of known high-latitude ($|b|>20 \degr$) clusters from the MWSC survey. We present the distributions of ``structural'' parameters, such as the total apparent radius $r_2$ of a cluster, the apparent radius $r_1$ of its densest central part, and the tidal radius $r_t$ derived by fitting a King profile to the observed distribution. We also show an empirical estimator of cluster richness $n_2$, i.\,e. the number of the most probable cluster members within $r_2$. The lower panel of Fig.~\ref{fig:cmprm} shows the distributions of the so-called  ``photometric'' parameters, derived from fitting cluster CMDs: age $\log t$, reddening $E(J-K_s)$, distance $\log d$, and the height $Z$ above the Galactic plane.

The data of the 139 new open clusters have been submitted to the CDS and the GAVO data center as an extension to the MWSC catalogue\footnote{ftp://cdsarc.u-strasbg.fr; http://dc.g-vo.org/mwsc-e14a/q/clu/form}. The format is the same as that of the MWSC survey in \citetalias{khea12}. An overview with positions, radii, distances and ages of the new clusters is given in Table~\ref{tbl:clu_list}.

\begin{figure}
\resizebox{\hsize}{!}{\includegraphics[clip=]{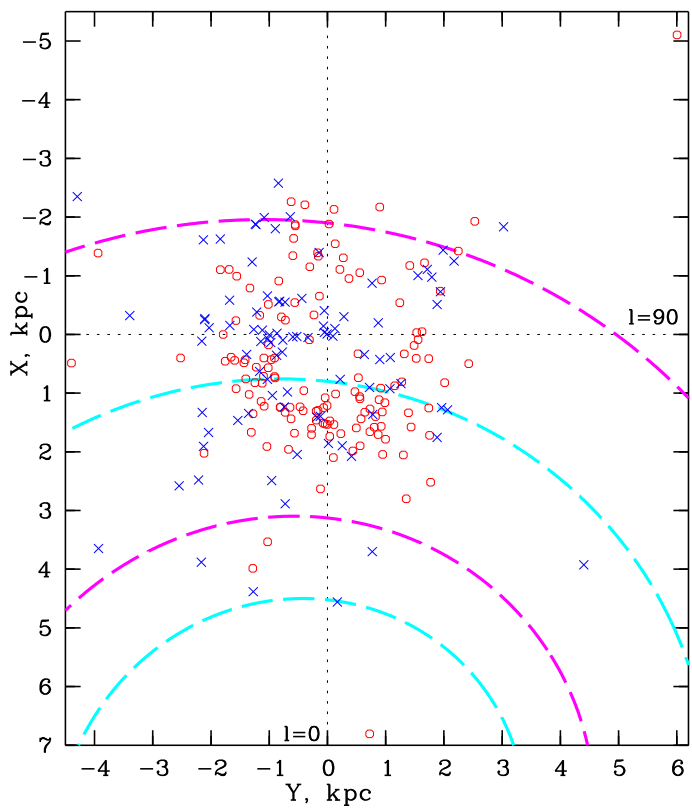}}
\caption{Distribution of the 139 new clusters (red circles) projected onto the Galactic $XY$-plane and of known open clusters (blue crosses) selected from the MWSC survey with $\log t=8.3\dots9.7$ and $|b|>20 \degr$. The dashed spirals indicate the positions of local spiral arms (magenta for Perseus and cyan for Sagittarius) as defined by the COCD clusters \citep{clupop}.}
\label{fig:xy_pos}
\end{figure}

\begin{figure}
\resizebox{\hsize}{!}{\includegraphics[clip=]{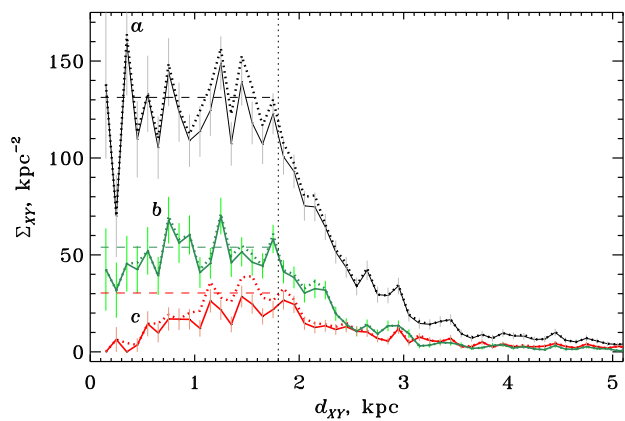}}
\caption{Contribution of the 139 new clusters to the surface density $\Sigma_{XY}$ versus the projected distance $d_{XY}$. The distribution of all clusters is given in black (a), and the distributions of two age groups are indicated with green ($\log t=8.3\dots9.0$, b) and red ($\log t>9.0$, c). Solid curves correspond to the densities of known open clusters from the MWSC survey, the dotted curves include the new clusters, the dotted vertical line marks the completeness limit found for the total sample, and the dashed horizontal lines correspond to the average surface density for different age groups.}
\label{fig:sden_pro}
\end{figure}

\section{Discussion}
\label{sec:discuss}

The initial goal of this search was to find unknown old star clusters at high Galactic latitudes, which might fill the local ``hole'' around the Sun, as we hoped. The results are illustrated in Figs.~\ref{fig:xy_pos} and \ref{fig:sden_pro} where we compare the distribution of known and new clusters in the $XY$-plane and show the contribution of new clusters to the surface density of Galactic open clusters.

\subsection{The ``hole'' around the Sun}

Figure~\ref{fig:xy_pos} shows that most of the newly discovered clusters occupy a ring around the Sun with inner and outer borders of $d_{XY} \approx$ 1 and 2 kpc ($d_{XY}$ is the cluster distance projected on the Galactic plane), with almost no clusters at $d_{XY} < $ 1 kpc. Figure~\ref{fig:sden_pro} indicates that the new clusters slightly increased (by about 8\%) the total surface density. The latter contribute mostly to the surface density of the oldest clusters ($\log t~\rm{[yr]}) >9.0$), which becomes higher and flatter within the ring. At $d_{XY} < $ 1 kpc, the shortage of the oldest clusters is now even more prominent. Assuming the average surface density within the ring to be typical for the whole range of the projected distances $d_{XY}$, we expect about 50 clusters still to be discovered in the solar vicinity. On the other hand, the new clusters do not significantly affect the surface density distribution of clusters with ages $8.3 < \log t < 9.0$, where a ``hole'' is only marginally visible at $d_{XY} \lesssim 0.5$ kpc. Possibly about ten clusters are missing in this age and distance range. There is no convincing reason for old clusters to avoid the area around the Sun, therefore it is more likely that they escaped our search because of its limitations as discussed below.

\subsection{Limitations of the search method}

Most likely, the missing clusters are just too sparse and too extended to be found as overdensities, even when applying our colour-magnitude filters. For example, it was not possible to detect the cluster \object{Ruprecht 147} ($d = 175$~pc, $\log t = 9.39$, $r_2 = 1\fdg23$; \citealt{clucat}) with our algorithm. In the area of Ruprecht 147 there are, even when applying our filters, more than 11\,000 field stars in 2MASS, compared to about 150 members found for this cluster in the MWSC survey. When only considering the cluster core, there are about 480 field stars compared to 20 cluster members. This is much smaller than the average noise. 
Even using a very narrow filter specifically tailored to the CMD of Ruprecht 147 instead of our standard filters does not reduce the background to a level where the cluster becomes detectable as an overdensity.
Similar to \citet{mercl05}, who added artificial clusters to their catalogue and tried to recover them, we did additional tests by simulating the \object{Hyades} ($d = 45$~pc, $t = 650$~Myr) at different distances between 0.6 and 2 kpc at a latitude of $b \approx 30 \degr$. It turns out that only at distances $\ge 1$~kpc is the innermost core ($r \approx 3$~pc) of the cluster detected as a significant overdensity.

\begin{figure}
\resizebox{\hsize}{!}{\includegraphics[clip=]{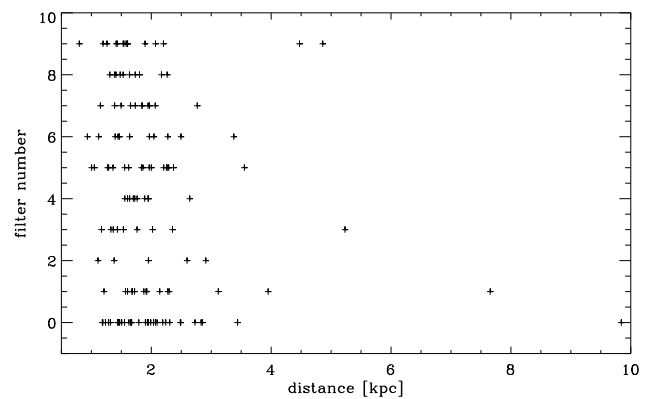}}
\caption{Filter number (0 = unfiltered search) versus distance of detected clusters. (Usually, a cluster is found in more than one filter, in those cases, the filter is considered where it shows the strongest signal.)}
\label{fig:dist_vs_filt}
\end{figure}

We also investigated the effect of our filters on the search.
A comparison of the results of both filtered and unfiltered searches (see Fig.~\ref{fig:dist_vs_filt},
showing the distances of new clusters identified with different filters) indicates 
that the distances do not strongly depend on a specific filter or on its absence.
Another experience we gained from the results of this search is that it seems that the presence of cluster 
members on the giant
branch facilitates their discovery with the filters, so an absence of giants may result in the clusters not being detected.

\begin{figure}
\resizebox{\hsize}{!}{\includegraphics[clip=]{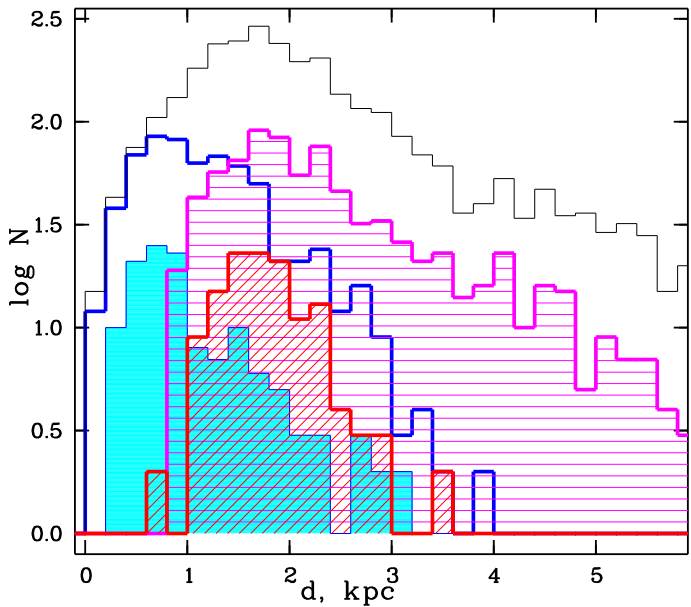}}
\caption{Comparison of distributions with distance of newly detected clusters for the optical COCD and NIR MWSC surveys. The  distributions of new clusters are shown with red (current sample), magenta \citep[candidates of][]{froeb07}, and cyan \citep[][for COCD]{newc109}. The total distributions are shown with black (MWSC) and blue (COCD).}
\label{fig:dst_cmp}
\end{figure}

To estimate the effect of the search method and the underlying catalogue, we compare
the distributions with distance of clusters detected in recent optical and NIR surveys (Fig.~\ref{fig:dst_cmp}). Both surveys differ by the basic catalogues they use and by the search algorithm. The optical data are represented by the Catalogue of Open Cluster Data \citep[COCD,][]{clucat,newc109}, based on the catalogue ASCC-2.5, which provides a higher accuracy of kinematic and photometric data and a lower level of background contamination than the combination PPMXL+2MASS does. Unlike the current detection algorithm, the new clusters in this study were searched as density enhancements in the four-dimensional space of proper motions and coordinates in the fields around bright stars ($V<9$ mag). In the case of MWSC, in addition to the current set of high latitude clusters we consider data on low latitude clusters of \citet{froeb07}, which are included in the MWSC input list.
While the total distribution of clusters in the NIR-based MWSC extends to higher distances than those of the optical survey COCD, their subsets of newly identified clusters differ with respect to the lower limit of their distances. While the bulk of new clusters found in the optical reside at distances less than 1~kpc, all the objects detected in the NIR are located outside the 1~kpc limit. This tendency is also seen in other detections of new objects based on the 2MASS catalogue \citep[see e.\,g.][]{gluea10}. One should note that all these results are based both on the same data source (2MASS) and use a similar approach of searching new clusters as density enhancements in the sky.

Other approaches, such as a search using proper motions \citep{nchpm}, may be more successful in finding the missing nearby clusters. In the long run, the {\em Gaia} mission is expected to fill the gap.

\section{Summary}

From a first-look analysis of the MWSC in \citetalias{mwscat}, we found evidence for a lack of nearby old clusters at high Galactic latitudes and projected distances $d_{XY} \lesssim1$ kpc. An additional search for star clusters was carried out on the basis of 2MASS and PPMXL at latitudes $|b| > 20 \degr$. We applied colour-magnitude filters and a star count algorithm to search for these old open clusters. This resulted in the detection of 782 overdensities, regarded as cluster candidates. A comparison with lists of known objects (MWSC input list, SIMBAD data base, and the list of Abell galaxy clusters) has shown that 383 of them are already known objects. The remaining 399 cluster candidates were processed with the standard MWSC pipeline, which confirmed the cluster nature of 139 objects. All of them are open clusters with ages $8.3 < \log t < 9.7$, distances $<3$\,kpc, and distances from the Galactic plane $0.3 < Z < 1$\,kpc. This increased the total number of known high-latitude open clusters by about 150\%.
Nevertheless, the ``hole'' with a radius of about 1\,kpc around the Sun could not be filled. This dearth of old clusters is expected to be an artefact from the bias against sparse overdensities with large angular size on the sky.
We estimate that still about 60 old open clusters are missing in this volume.

\begin{acknowledgements}
We wish to thank the referee for a detailed and helpful report.
This study was supported by Sonderforschungsbereich SFB 881 ``The Milky Way System'' (subproject B5) of the German Research Foundation (DFG) and by DFG grant RO~528/10-1. This publication makes use of data products from the Two Micron All Sky Survey, which is a joint project of the University of Massachusetts and the Infrared Processing and Analysis Center/California Institute of Technology, funded by the National Aeronautics and Space Administration and the National Science Foundation. This research has made use of the SIMBAD database, operated at the CDS, Strasbourg, and of the WEBDA database, operated at the Institute for Astronomy of the University of Vienna.
\end{acknowledgements}

\bibliographystyle{aa}
\bibliography{hlc_arxiv}

\Online
\begin{appendix} 

\section{Atlas page of MWSC 5244}

\begin{figure}[t]
\resizebox{\hsize}{!}{
\includegraphics[origin=l,angle=270,clip=]{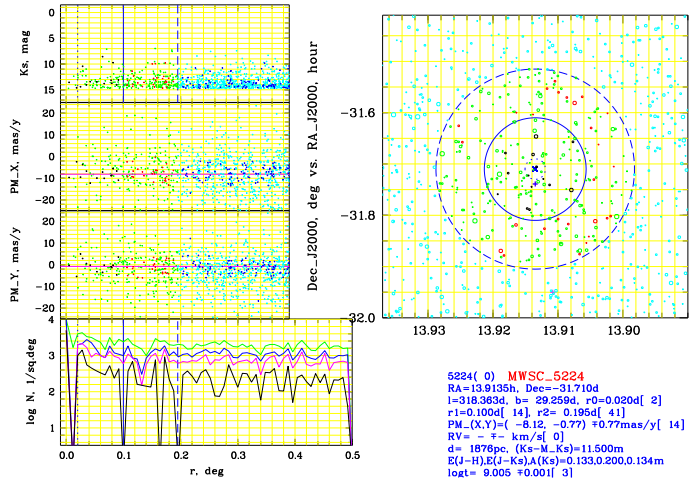}
}
\caption{Newly found cluster MWSC 5224 in the MWSC Atlas (page~1). 
Sky map of the cluster region with the most probable members shown in black and red and the
radii $r_0$, $r_1$, and $r_2$ as dotted, solid, and dashed line, respectively (right panel); 
$K_S$ magnitude and proper motions versus cluster radius (upper left panels) 
and radial density profile (lower left panel).
See text for a detailed explanation.
}
\label{fig:atlp1}
\end{figure}

Figures~\ref{fig:atlp1} and \ref{fig:atlp2} show the pipeline output for the newly identified cluster MWSC 5224 to illustrate the star member selection procedure and the quality of the determined cluster parameters. The example cluster is selected at random and represents a typical case among the analysed objects.

The main diagram of the first page of the Atlas is a cluster map, while on the second page this role is played by the $K_s,(J-H)$, and $K_s,(J-K_s)$ diagrams. Stars are shown as coloured circles or dots. Symbols and their colours have the same meaning in all plots. Cyan symbols mark stars outside the cluster radius $r_2$ and green symbols stars within $r_2$. The most probable kinematic and photometric members ($1\sigma$-members) are indicated in black for members located within $r_1$, red for members between $r_1$ and $r_2$, and blue for stars outside $r_2$. Cyan bars show the uncertainty for $1\sigma$-members (page 2).

Page~1 of the Atlas (Fig.~\ref{fig:atlp1}) contains five diagrams with spatial information, as well as a legend on the derived cluster parameters. The right panel is a map of the cluster surroundings. The left panels show magnitudes $K_s$, proper motions $PM_x$, $PM_y$, and surface density $N$ versus distances $r$ of stars from the cluster centre.

In the sky map, stars are shown by circles. Their size corresponds to the brightness arranged in six $K_s$ magnitude bins. The blue cross indicates the cluster centre determined in this study. If by chance other clusters appear in this area, their centres are marked by magenta plus signs. Large blue circles (shown by dotted, solid, or dashed curves) indicate the cluster radii $r_0$, $r_1$, and $r_2$, respectively. In the left-hand panels, the blue vertical lines (dotted, solid, or dashed) mark $r_0$, $r_1$, or $r_2$. Magenta horizontal lines in the $PM$ vs. $r$ diagrams correspond to the derived average proper motion of the cluster. Radial density profiles in the bottom panel are shown with green for all stars, blue for $3\sigma$-members, magenta for $2\sigma$-members, and black for $1\sigma$-members.

\begin{figure}[t]
\resizebox{\hsize}{!}{
\includegraphics[origin=l,angle=270,clip=]{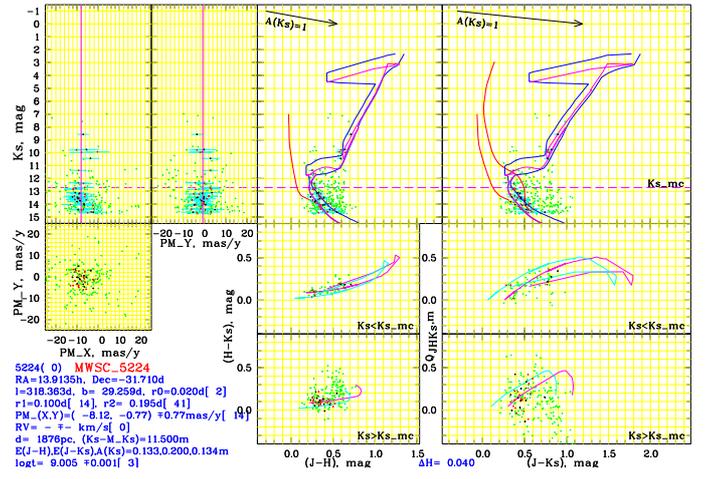}
}
\caption{Newly found cluster MWSC 5224 in the MWSC Atlas (page~2):
Proper motion relations (left panel), CMDs (upper right panels), two-colour diagrams (lower central panels)
and $Q_{JHK}$-colour diagrams (lower right panels).
See text for a detailed explanation.
}
\label{fig:atlp2}
\end{figure}

The legend gives cluster name, MWSC number, and COCD number in parentheses; equatorial $RA_{J2000}$, $Dec_{J2000}$, and Galactic $l,\,b$ coordinates of the cluster centre; apparent cluster sizes $r_0$, $r_1$, $r_2$, and number of $1\sigma$-members within the corresponding radius; weighted average components $PM_{X,Y}$ of proper motion with their $rms$ errors and number of stars used to compute the average; the average radial velocity, $RV$, $rms$ error, and the number of stars used to compute the average; distance to the cluster, $d$, distance modulus, $(K_s-M_{K_s})$; NIR interstellar reddening, $E(J-H),\,E(J-Ks)$, and interstellar extinction, $A(K_s)$; cluster age, and its \textit{rms} error. The number in brackets gives the number of stars used to compute the average age, or it is -1 if an isochrone fitting was applied. The $\Delta H$ shown below the photometric diagrams indicates the empirical correction to the $H$-magnitude introduced in \citetalias{khea12}.

The parameters are shown as they were derived in the pipeline without taking their real accuracy into account, which was estimated by us from comparison with literature data after the MWSC was completed (for details see \citetalias{mwscat}).
Typically the cluster proper motions are accurate within 1 mas/yr, and the derived distances and reddenings are accurate within 11\% and 7\% respectively. An accuracy of the order of 10\% is achieved for the ages of older open clusters ($\log (t~\rm{[yr]}) > 8.2$).

Page 2 (Fig.~\ref{fig:atlp2}) contains three diagrams with kinematic information (left panels), and six diagrams with photometric information (right panels).

The three left panels with kinematic data: the two upper diagrams show $PM_{X,Y}$ vs. $K_s$ relations, i.e.``PM-magnitude equation''. Magenta vertical lines correspond to the average proper motion of the cluster. The magenta dashed line shows the apparent magnitude $K_s^{mc}$, which corresponds to the bluest colour $(J-K_s)$ of the adopted isochrone. The bottom panel is the vector point diagram of proper motions.

The six right panels with photometric data: the two upper diagrams are CMDs ($K_s,(J-H)$ and $K_s,(J-K_s)$). The magenta curve is the apparent isochrone closest to the determined cluster age. Solid blue lines outline a domain of 100\% photometric members. Solid red lines (shown only in $K_s,(J-K_s)$,) are the ZAMS (zero-age main sequence) and TAMS (terminal-age main sequence), described in more detail in \citetalias{khea12}. The magenta dashed line shows the apparent magnitude of minimum colour $K_s^{mc}$. The thick yellow circles mark the stars used for the age determination \citep[see][for details]{clucat}. The black arrows show the vectors of increasing extinction.   The four bottom panels show the two-colour $(H-K_s)/(J-H)$ diagram (left column) and $Q_{JHK}$-colour diagram (right column). The upper row is for stars brighter than $K_s^{mc}$, the lower row is for stars fainter than $K_s^{mc}$. Magenta curves indicate the apparent isochrone (i.e., apparent colours), whereas cyan curves show the intrinsic isochrone.

The legend is the same as on Page 1.

The atlas pages for all new clusters will be available in electronic form at the CDS and at the GAVO data center\footnote{http://dc.g-vo.org/mwsc-e14a/q/clu/form}.

\section{Table of newly identified clusters}

Table~\ref{tbl:clu_list} gives an overview of the 139 newly identified clusters. Since the total list of determined parameters is too long (37 columns), we only show the most important parameters here for quick reference (cluster names, equatorial and Galactic coordinates, their total sizes $r_2$, distance, and age). The full list of cluster parameters is available in electronic form at the CDS. It is in the same format as the table determined earlier in Paper~II for the main body of the MWSC survey.

\begin{longtable}{l r r r r r r r}
\caption{\label{tbl:clu_list}The list of newly discovered high-latitude MWSC clusters}\\
\hline\hline
 Name      & RA    & Dec     & $l$     & $b$      &$r_2$     & distance & age\\
           & {\footnotesize [hr] (J2000)} & {\footnotesize [deg] (J2000)} & {\footnotesize [deg]} & {\footnotesize [deg]} & {\footnotesize [deg]}    & {\footnotesize [pc]} & {\footnotesize $\log(t~\rm{[yr])}$}\\
\hline
\endfirsthead
\caption{continued.}\\
\hline\hline
 Name      & RA    & Dec     & $l$     & $b$      &$r_2$     & distance & age\\
           & {\footnotesize [hr]}  & {\footnotesize [deg]}     & {\footnotesize [deg]}   & {\footnotesize [deg]}    & {\footnotesize [deg]}    & {\footnotesize [pc]}     &{\footnotesize $\log(t~\rm{[yr])}$}\\
\hline
\endhead
\hline
\endfoot
MWSC\_5004 &   4.298 &  86.183 & 126.223 &  24.716 & 0.170 & 2274 & 9.215 \\
MWSC\_5010 &   6.305 &  86.035 & 127.314 &  26.485 & 0.130 & 3553 & 9.450 \\
MWSC\_5011 &   6.410 &  62.630 & 152.192 &  20.919 & 0.155 & 1272 & 9.450 \\
MWSC\_5012 &   6.456 &  57.460 & 157.507 &  19.514 & 0.125 & 2492 & 9.085 \\
MWSC\_5016 &   7.154 &  40.420 & 176.994 &  20.373 & 0.150 & 2276 & 8.810 \\
MWSC\_5018 &   7.575 &  22.825 & 196.512 &  19.269 & 0.145 & 2070 & 9.120 \\
MWSC\_5019 &   7.676 &  32.780 & 186.948 &  23.949 & 0.150 & 1534 & 8.970 \\
MWSC\_5022 &   8.273 &   0.450 & 222.559 &  19.021 & 0.130 & 1308 & 9.170 \\
MWSC\_5029 &   8.760 & $-$11.655 & 237.489 &  18.928 & 0.135 & 1953 & 9.285 \\
MWSC\_5033 &   8.932 &  30.155 & 194.733 &  38.814 & 0.205 & 1530 & 9.170 \\
MWSC\_5038 &   9.238 &  23.940 & 203.901 &  41.284 & 0.125 & 1953 & 9.300 \\
MWSC\_5042 &   9.351 &  $-$7.475 & 239.297 &  28.473 & 0.200 & 1762 & 9.500 \\
MWSC\_5044 &   9.419 &  29.925 & 196.639 &  44.974 & 0.140 & 2729 & 9.570 \\
MWSC\_5051 &   9.764 &  $-$7.005 & 243.267 &  33.615 & 0.195 & 1365 & 9.450 \\
MWSC\_5058 &  10.138 &  12.318 & 225.930 &  49.075 & 0.230 & 1191 & 8.950 \\
MWSC\_5060 &  10.175 & $-$14.045 & 254.381 &  33.254 & 0.190 & 1445 & 9.700 \\
MWSC\_5062 &  10.206 &  $-$9.181 & 250.618 &  37.050 & 0.120 & 5232 & 9.450 \\
MWSC\_5071 &  10.545 & $-$29.585 & 270.048 &  24.290 & 0.150 & 1965 & 9.450 \\
MWSC\_5076 &  10.778 &  78.450 & 130.371 &  36.809 & 0.120 & 9842 & 8.850 \\
MWSC\_5083 &  10.933 & $-$32.405 & 276.330 &  24.437 & 0.125 & 4860 & 9.200 \\
MWSC\_5088 &  11.146 & $-$32.530 & 279.044 &  25.537 & 0.185 & 2831 & 9.225 \\
MWSC\_5116 &  11.651 & $-$36.340 & 287.001 &  24.286 & 0.180 & 1622 & 9.515 \\
MWSC\_5117 &  11.672 & $-$17.835 & 279.955 &  41.844 & 0.190 & 1380 & 9.360 \\
MWSC\_5122 &  11.732 & $-$28.798 & 285.506 &  31.781 & 0.170 & 1947 & 9.500 \\
MWSC\_5149 &  12.280 & $-$32.990 & 294.589 &  29.322 & 0.130 & 1542 & 9.500 \\
MWSC\_5154 &  12.416 &   4.045 & 286.447 &  66.068 & 0.260 &  799 & 9.500 \\
MWSC\_5186 &  13.155 & $-$32.970 & 307.232 &  29.756 & 0.205 & 1947 & 9.325 \\
MWSC\_5191 &  13.241 & $-$34.190 & 308.331 &  28.444 & 0.161 & 1318 & 8.975 \\
MWSC\_5215 &  13.767 & $-$41.843 & 313.703 &  19.887 & 0.175 & 3118 & 9.290 \\
MWSC\_5224 &  13.913 & $-$31.710 & 318.363 &  29.259 & 0.195 & 1876 & 9.005 \\
MWSC\_5231 &  14.142 & $-$19.465 & 326.818 &  39.809 & 0.160 & 1935 & 9.315 \\
MWSC\_5273 &  15.118 & $-$36.361 & 331.281 &  18.940 & 0.165 & 1599 & 9.445 \\
MWSC\_5279 &  15.193 & $-$21.428 & 341.397 &  30.783 & 0.200 & 2066 & 9.100 \\
MWSC\_5289 &  15.389 & $-$32.490 & 336.519 &  20.302 & 0.185 & 1672 & 9.450 \\
MWSC\_5292 &  15.453 & $-$26.487 & 341.147 &  24.600 & 0.150 & 2276 & 9.365 \\
MWSC\_5293 &  15.505 & $-$30.440 & 339.100 &  21.061 & 0.165 & 1414 & 9.220 \\
MWSC\_5295 &  15.578 & $-$13.380 & 352.490 &  33.365 & 0.170 & 1764 & 9.360 \\
MWSC\_5299 &  15.698 & $-$29.000 & 342.166 &  20.604 & 0.130 & 4473 & 9.400 \\
MWSC\_5300 &  15.729 & $-$10.300 & 357.003 &  33.930 & 0.195 & 1506 & 9.400 \\
MWSC\_5301 &  15.752 & $-$27.510 & 343.789 &  21.269 & 0.150 & 3950 & 9.300 \\
MWSC\_5309 &  16.161 & $-$24.440 & 350.308 &  19.698 & 0.185 & 1724 & 9.025 \\
MWSC\_5311 &  16.194 & $-$23.215 & 351.580 &  20.219 & 0.155 & 1410 & 9.050 \\
MWSC\_5312 &  16.193 & $-$15.960 & 357.392 &  25.109 & 0.170 & 2910 & 9.100 \\
MWSC\_5316 &  16.254 & $-$22.370 & 352.845 &  20.185 & 0.210 & 1390 & 8.870 \\
MWSC\_5318 &  16.345 & $-$17.235 & 357.893 &  22.621 & 0.150 & 1169 & 9.585 \\
MWSC\_5319 &  16.350 & $-$15.143 & 359.690 &  23.908 & 0.165 & 1326 & 9.500 \\
MWSC\_5321 &  16.447 &  $-$8.927 &   6.076 &  26.616 & 0.140 & 7655 & 9.270 \\
MWSC\_5323 &  16.451 &  $-$7.170 &   7.707 &  27.606 & 0.170 & 1925 & 9.375 \\
MWSC\_5326 &  16.546 & $-$17.143 & 359.941 &  20.449 & 0.180 & 1637 & 9.485 \\
MWSC\_5329 &  16.718 &  $-$9.840 &   7.837 &  22.837 & 0.175 & 1111 & 9.650 \\
MWSC\_5333 &  16.815 &  16.900 &  35.462 &  34.510 & 0.170 & 1968 & 9.350 \\
MWSC\_5337 &  17.124 &  $-$3.335 &  17.276 &  21.294 & 0.125 & 1791 & 9.390 \\
MWSC\_5338 &  17.130 &  12.105 &  32.395 &  28.423 & 0.165 & 2767 & 9.405 \\
MWSC\_5340 &  17.174 &   6.425 &  26.981 &  25.357 & 0.170 & 1361 & 9.465 \\
MWSC\_5343 &  17.240 &  17.300 &  38.555 &  29.006 & 0.135 & 1626 & 9.355 \\
MWSC\_5344 &  17.258 &   4.570 &  25.800 &  23.398 & 0.185 & 1604 & 9.320 \\
MWSC\_5346 &  17.287 &   2.212 &  23.764 &  21.910 & 0.180 & 1952 & 9.310 \\
MWSC\_5348 &  17.319 &  13.505 &  35.107 &  26.477 & 0.110 & 3440 & 9.500 \\
MWSC\_5350 &  17.367 &   4.915 &  26.947 &  22.112 & 0.145 & 1914 & 9.475 \\
MWSC\_5351 &  17.387 &   5.342 &  27.505 &  22.041 & 0.120 & 1288 & 9.200 \\
MWSC\_5354 &  17.423 &  13.465 &  35.751 &  25.067 & 0.210 & 1004 & 9.250 \\
MWSC\_5356 &  17.473 &   2.890 &  25.813 &  19.766 & 0.180 & 1151 & 9.205 \\
MWSC\_5358 &  17.619 &   6.135 &  29.937 &  19.304 & 0.150 & 1553 & 9.070 \\
MWSC\_5359 &  17.635 &   6.618 &  30.508 &  19.307 & 0.130 & 1721 & 9.205 \\
MWSC\_5365 &  17.904 &  28.850 &  54.222 &  24.369 & 0.130 & 2354 & 8.985 \\
MWSC\_5366 &  17.986 &  31.830 &  57.691 &  24.307 & 0.145 & 1658 & 9.285 \\
MWSC\_5367 &  18.053 &  20.102 &  46.186 &  19.361 & 0.155 & 2044 & 9.150 \\
MWSC\_5368 &  18.058 &  19.447 &  45.574 &  19.046 & 0.105 & 2597 & 9.255 \\
MWSC\_5370 &  18.721 &  46.775 &  75.889 &  20.739 & 0.135 & 1454 & 9.350 \\
MWSC\_5371 &  18.809 &  47.235 &  76.657 &  20.030 & 0.140 & 1902 & 9.035 \\
MWSC\_5373 &  19.090 &  56.340 &  86.739 &  20.457 & 0.130 & 1664 & 9.100 \\
MWSC\_5374 &  19.305 &  60.640 &  91.659 &  20.270 & 0.175 & 1732 & 9.260 \\
MWSC\_5377 &  21.559 &  78.256 & 113.502 &  19.189 & 0.160 & 1434 & 9.245 \\
MWSC\_5901 &   7.504 &  27.020 & 191.995 &  19.916 & 0.240 &  709 & 9.380 \\
\hline
MWSC\_5533 &   0.807 &  41.600 & 122.314 & $-$21.271 & 0.160 & 2853 & 8.800 \\
MWSC\_5558 &   1.378 &  37.965 & 129.693 & $-$24.499 & 0.160 & 2020 & 9.125 \\
MWSC\_5571 &   1.862 &  41.500 & 134.889 & $-$19.971 & 0.155 & 1397 & 9.310 \\
MWSC\_5572 &   1.905 & $-$78.965 & 299.175 & $-$37.698 & 0.140 & 1472 & 9.300 \\
MWSC\_5575 &   1.995 & $-$83.050 & 300.483 & $-$33.748 & 0.150 & 2191 & 9.200 \\
MWSC\_5602 &   3.082 &  23.095 & 158.811 & $-$30.325 & 0.155 & 1184 & 9.390 \\
MWSC\_5604 &   3.164 & $-$42.880 & 251.404 & $-$57.905 & 0.200 & 1437 & 9.200 \\
MWSC\_5621 &   3.810 &  22.270 & 168.239 & $-$24.630 & 0.140 & 1468 & 9.460 \\
MWSC\_5623 &   4.019 &  24.115 & 169.098 & $-$21.307 & 0.185 & 1211 & 9.100 \\
MWSC\_5627 &   4.319 &  20.335 & 175.113 & $-$20.867 & 0.165 & 1659 & 9.250 \\
MWSC\_5633 &   4.557 &   9.400 & 186.720 & $-$25.025 & 0.161 & 1481 & 9.100 \\
MWSC\_5634 &   4.589 &  18.540 & 179.183 & $-$19.105 & 0.155 & 1991 & 9.250 \\
MWSC\_5645 &   5.015 &  10.445 & 189.950 & $-$18.875 & 0.180 & 2371 & 9.010 \\
MWSC\_5651 &   5.179 &   5.890 & 195.402 & $-$19.263 & 0.165 & 2485 & 9.225 \\
MWSC\_5656 &   5.278 &   2.210 & 199.560 & $-$19.873 & 0.195 & 1847 & 8.800 \\
MWSC\_5665 &   6.174 & $-$42.210 & 249.354 & $-$25.068 & 0.140 &  934 & 8.875 \\
MWSC\_5667 &   6.300 & $-$31.800 & 239.068 & $-$20.506 & 0.180 & 2290 & 9.015 \\
MWSC\_5668 &   6.326 & $-$29.365 & 236.763 & $-$19.348 & 0.180 & 2141 & 8.900 \\
MWSC\_5670 &   7.057 & $-$51.205 & 261.485 & $-$19.043 & 0.125 & 1682 & 8.950 \\
MWSC\_5671 &   7.085 & $-$73.380 & 284.576 & $-$24.945 & 0.210 & 2001 & 9.400 \\
MWSC\_5672 &   7.255 & $-$78.440 & 290.189 & $-$25.350 & 0.135 & 1288 & 9.490 \\
MWSC\_5674 &   8.109 & $-$70.325 & 283.172 & $-$19.443 & 0.115 & 1803 & 9.350 \\
MWSC\_5676 &   8.698 & $-$74.765 & 288.641 & $-$19.361 & 0.145 & 1604 & 9.150 \\
MWSC\_5679 &   9.550 & $-$78.970 & 294.025 & $-$19.705 & 0.155 & 1122 & 8.500 \\
MWSC\_5680 &   9.730 & $-$78.205 & 293.857 & $-$18.786 & 0.150 & 1050 & 9.300 \\
MWSC\_5681 &  10.696 & $-$82.005 & 298.370 & $-$20.304 & 0.115 & 1700 & 8.980 \\
MWSC\_5684 &  12.895 & $-$86.648 & 302.966 & $-$23.773 & 0.155 & 1432 & 9.180 \\
MWSC\_5685 &  13.090 & $-$82.043 & 303.440 & $-$19.182 & 0.160 & 1581 & 9.150 \\
MWSC\_5688 &  15.433 & $-$80.130 & 309.439 & $-$19.269 & 0.105 & 1236 & 8.990 \\
MWSC\_5691 &  17.436 & $-$70.735 & 321.885 & $-$18.881 & 0.100 & 2241 & 8.350 \\
MWSC\_5692 &  17.789 & $-$86.610 & 306.560 & $-$26.142 & 0.135 & 1555 & 8.930 \\
MWSC\_5694 &  17.956 & $-$66.800 & 326.935 & $-$19.698 & 0.110 & 1570 & 9.000 \\
MWSC\_5696 &  18.188 & $-$62.815 & 331.444 & $-$19.549 & 0.120 & 2308 & 8.550 \\
MWSC\_5697 &  18.410 & $-$62.195 & 332.632 & $-$20.778 & 0.125 & 1496 & 9.250 \\
MWSC\_5698 &  18.697 & $-$77.723 & 316.596 & $-$25.947 & 0.150 & 2069 & 8.890 \\
MWSC\_5701 &  18.826 & $-$53.860 & 342.212 & $-$21.337 & 0.130 & 1467 & 8.750 \\
MWSC\_5704 &  19.073 & $-$39.930 & 357.158 & $-$19.329 & 0.110 & 1604 & 9.370 \\
MWSC\_5705 &  19.115 & $-$46.272 & 350.885 & $-$21.768 & 0.120 & 1861 & 8.950 \\
MWSC\_5706 &  19.160 & $-$35.860 &   1.583 & $-$18.918 & 0.140 & 1558 & 8.990 \\
MWSC\_5708 &  19.199 & $-$36.315 &   1.303 & $-$19.518 & 0.100 & 1844 & 8.575 \\
MWSC\_5712 &  19.422 & $-$34.475 &   4.119 & $-$21.474 & 0.140 & 1654 & 8.745 \\
MWSC\_5713 &  19.426 & $-$35.820 &   2.762 & $-$21.945 & 0.095 & 2265 & 9.115 \\
MWSC\_5715 &  19.559 & $-$26.895 &  12.402 & $-$20.564 & 0.150 & 2170 & 8.360 \\
MWSC\_5717 &  19.578 & $-$22.888 &  16.469 & $-$19.336 & 0.105 & 2097 & 8.775 \\
MWSC\_5720 &  19.695 & $-$18.105 &  21.810 & $-$19.004 & 0.110 & 1639 & 9.360 \\
MWSC\_5723 &  19.701 & $-$60.015 & 337.003 & $-$29.497 & 0.150 & 1195 & 9.130 \\
MWSC\_5726 &  19.890 & $-$13.960 &  27.037 & $-$19.919 & 0.135 & 2038 & 8.900 \\
MWSC\_5731 &  20.047 & $-$16.220 &  25.792 & $-$22.905 & 0.155 & 3377 & 9.265 \\
MWSC\_5732 &  20.069 & $-$12.928 &  29.186 & $-$21.863 & 0.160 & 2203 & 8.650 \\
MWSC\_5735 &  20.156 & $-$12.185 &  30.489 & $-$22.722 & 0.125 & 1970 & 9.250 \\
MWSC\_5737 &  20.184 & $-$17.940 &  24.876 & $-$25.370 & 0.135 & 2495 & 8.925 \\
MWSC\_5740 &  20.262 &  $-$0.630 &  42.304 & $-$18.954 & 0.135 & 2255 & 8.825 \\
MWSC\_5744 &  20.339 & $-$42.245 & 358.445 & $-$33.757 & 0.115 & 1836 & 9.220 \\
MWSC\_5745 &  20.351 &  $-$3.530 &  40.258 & $-$21.511 & 0.140 & 1639 & 9.080 \\
MWSC\_5748 &  20.483 &   0.400 &  44.977 & $-$21.355 & 0.150 & 1388 & 8.725 \\
MWSC\_5749 &  20.536 & $-$78.615 & 315.018 & $-$31.446 & 0.150 & 1890 & 9.345 \\
MWSC\_5751 &  20.596 &   7.985 &  52.837 & $-$18.912 & 0.140 & 1532 & 9.250 \\
MWSC\_5764 &  21.160 &  19.486 &  67.765 & $-$18.873 & 0.120 & 2297 & 9.200 \\
MWSC\_5779 &  21.729 &  25.820 &  78.371 & $-$20.392 & 0.135 & 2642 & 9.130 \\
MWSC\_5782 &  21.767 &  20.830 &  75.002 & $-$24.327 & 0.150 & 1732 & 9.290 \\
MWSC\_5800 &  22.606 &  30.240 &  91.168 & $-$24.177 & 0.160 & 1679 & 8.325 \\
MWSC\_5804 &  22.732 &  15.040 &  82.777 & $-$37.695 & 0.200 & 1896 & 9.465 \\
MWSC\_5811 &  23.051 & $-$12.315 &  57.842 & $-$60.611 & 0.240 & 1261 & 9.425 \\
MWSC\_5828 &  23.845 &  41.428 & 110.860 & $-$20.019 & 0.180 & 2207 & 9.200 \\
MWSC\_5963 &   5.732 & $-$10.655 & 215.029 & $-$19.770 & 0.250 &  430 & 8.820 \\
\end{longtable}

\end{appendix}

\end{document}